\documentclass[aps,preprint]{revtex4}

\begin{document}

\title{On black strings \& branes in Lovelock gravity}
\author{David Kastor$^1$\footnote{email address:
kastor@physics.umass.edu} and Robert Mann$^{2,3}$\footnote{email address:
rbmann@sciborg.uwaterloo.ca}}

\address{$^1$Department of Physics, University of Massachusetts, Amherst, MA 01003\\
$^2$Department of Physics, University of Waterloo, 200 University
Avenue West, Waterloo, Ontario,
Canada, N2L 3G1 \\
$^3$Perimeter Institute for Theoretical Physics, 35 Caroline St.
N., Waterloo, Ont. Canada}

\begin{abstract}
It is well known that black strings and branes may be constructed
in pure Einstein gravity simply by adding flat directions to a
vacuum black hole solution.   A similar construction holds in the
presence of a cosmological constant.  While these constructions
fail in general Lovelock theories, we show that they carry over
straightforwardly within a class of Lovelock gravity theories that
have (locally) unique constant curvature vacua.
\end{abstract}
\maketitle

\section{Introduction}

It is straightforward to construct uncharged black strings and branes in
Einstein gravity. Starting from a Schwarzschild black hole in $D$
-dimensions, for example, one obtains a black string in $(D+1)$-dimensions
simply by adding a flat direction to the metric. Despite the seemingly
trivial nature of this construction, the phenomenology associated with black
strings, begining with the Gregory-Laflamme instability \cite{Gregory:1993vy}
, has turned out to be surprisingly rich (see reference \cite{Kol:2004ww}
for a review). One can similarly construct black $p$-branes by adding
additional flat directions, while rotating black branes can be obtained by
adding flat directions to the $D$-dimensional Myers-Perry metrics \cite{Myers:1986un}.

Lovelock gravity theories \cite{Lovelock:1971yv} are fascinating extensions
of general relativity that include higher curvature interactions. The
Lagrangian density for Lovelock gravity in $D$ spacetime dimensions can be
written $\mathcal{L}=\sum_{k=0}^{[d/2]}c_k \mathcal{L}_k$, where
\begin{equation}
\mathcal{L}_k= {\frac{1}{2^k}}\sqrt{-g}\, \delta^{a_1\dots a_{k}b_1\dots b_k}_{c_1\dots c_{k}d_1\dots d_k} \,
R_{a_1b_1} {}^{c_1d_1}\dots R_{a_kb_k} {}^{c_kd_k},
\end{equation}
and $\mathcal{L}_0=\sqrt{-g}$. The $\delta$ symbol above denotes the totally
antisymmetric product of $2k$ Kronecker deltas, normalized to take values $0$
and $\pm 1$. The term $\mathcal{L}_0$ gives the cosmological term in the
action, while $\mathcal{L}_1=\sqrt{-g}R$ gives the Einstein term. The
curvature squared term, known as the Gauss-Bonnet term, may be expanded to
give $\mathcal{L}_2=\sqrt{-g}(R_{abcd}R^{abcd}-4R_{ab}R^{ab}+R^2)$. In $D=4$
this term can be written as a divergence and does not contribute to the
equations of motion. Similarly the term $\mathcal{L}_k$ is the
Euler density in $D=2k$ and hence contributes to the equations of motion only for dimensions $D>2k$. 
Lovelock theories are distinguished, among the much larger class of general
higher curvature theories, by having field equations involving not more than
second derivatives of the metric. Consequently, Lovelock gravity theories
are free from many of the pathologies that plague general higher derivative
gravity theories.

The black hole solutions of Lovelock gravity have been been studied begining with
the work of  \cite{Boulware:1985wk}\cite{Wheeler:1985nh}\cite{Wheeler:1985qd} some $20$ years ago.
One would also like to know the black string and black brane solutions,
\textit{e.g.} in order to study the effects of higher curvature interactions
on black string phenomenology. A number of workers have tried to find such
solutions. However, one immediately
encounters the fact that adding flat directions does not work in the
general Lovelock theory.  This was noted for Gauss-Bonnet gravity in reference
\cite{Barcelo:2002wz}.  Thus far, analytic solutions for black branes in general Lovelock theories
have not been found, and workers have turned to other methods.
In particular, black strings in $5$-dimensional Gauss-Bonnet gravity
were studied numerically in reference \cite{Kobayashi:2004hq} and more general black branes
in this theory were studied via near horizon and far field
expansions in reference \cite{Sahabandu:2005ma}.

In this paper, we show that black brane solutions to Lovelock gravity theories
including higher curvature terms may, in fact, be
simply constructed, but only within a certain class of Lovelock theories.
This class of theories has the following
property.  Assume that $\mathcal{L}_{p}$ is the highest order term in the
Lagrangian, \textit{i.e.} that the coefficients $c_{k}$ vanish for $k>p$. Depending on the values of
the nonzero coefficients in the Lagrangian, it then turns out that the theory may have up to $p$
distinct constant curvature vacuum solutions  \cite{Boulware:1985wk}\cite{Wheeler:1985nh}\cite{Wheeler:1985qd}.
The different values that the constant curvature may take are
the roots of a $p$th order polynomial.  There will, of course, generally be $p$ roots, but
only real values of  the curvature are considered to be physical.
The coefficients in the Lovelock Lagrangian may be tuned such that there are $p$ real roots and that all these roots coincide. The
theory then  has a (locally) unique constant curvature vacuum solution. We will refer to these
as LUV theories - standing for Lovelock-Unique-Vacuum.   We show that LUV theories have simple black brane solutions.

The simplest LUV theory with non-trivial dynamics is
Einstein gravity with a cosmological constant $\Lambda$, and our
results on brane solutions will be, roughly, that what works in
this theory works in all LUV theories.   We can first consider
Einstein gravity with $\Lambda=0$. In this case, as described
above, brane solutions are simply obtained by adding flat
directions.   Our first result will be that adding flat directions
also works in LUV theories, in the limit that the curvature of the
vacuum has been tuned to zero. Taking this limit of a LUV theory
with highest order interaction ${\cal L}_p$ sends the coefficients
of all the lower order terms to zero and we have simply ${\cal
L}={\cal L}_p$.  We will call these pure Lovelock gravity
theories, with pure Einstein gravity as the first non-trivial
example.

If we now consider Einstein gravity with $\Lambda\ne 0$,
additional directions may still be added, but they are no longer
flat, {\it i.e.} the new metric must have nontrivial coordinate
dependence on the new direction.  A good example of this is the
AdS black string of reference \cite{Chamblin:1999by}, which is
given by
\begin{equation}\label{chrstring}
ds^2 = {l^2\over z^2} \left (dz^2 - (1- {2m\over r}) dt^2 + {dr^2\over (1- {2m\over r})}+ r^2  d\Omega^2\right )
\end{equation}
where $l$ is the radius of curvature of $5$-dimensional AdS.  This
differs from the black string of pure Einstein gravity in the
overall conformal rescaling by the function $l^2/z^2$ of the new
coordinate $z$.  Note also that the $4$-dimensional `seed'  metric
for the construction, the Schwarzschild metric in this case, is a
solution to Einstein gravity with vanishing cosmological constant.
More generally the $4$-dimensional Schwarzschild metric in
(\ref{chrstring}) may be replaced by any Ricci flat metric.  We
will show below that in LUV theories with nonzero vacuum
curvature, new directions may be added in a similar way.  The
metric will generally be conformally scaled by a function of the
new coordinate.  The lower dimensional seed metric can  be a
solution to a LUV theory with a different value of the vacuum
curvature, possibly zero as in the metric (\ref{chrstring}) above.

LUV theories and their black hole solutions have been extensively
discussed in reference \cite{Crisostomo:2000bb}.   It is argued there that
LUV theories have a number of nice properties that distinguish them as possiby the
most physically relevant Lovelock theories.  We note \cite{Crisostomo:2000bb} that
in odd dimensions $D=2p+1$ the LUV  theory with highest term $\mathcal{L}_p$, which is the
highest non-trivial term,
corresponds to Chern-Simons gravity\footnote{Further properties of these theories are discussed in references \cite{Mora:2004kb}\cite{Banados:2005rz}.}, while in even dimensions the LUV theory that includes the highest non-trivial interaction follows from a Born-Infeld type action.

\section{LUV Theories}


In this section we recount some useful results in Lovelock gravity, focusing in particular on the LUV theories.
The equations of motion following from the Lovelock Lagrangian
$\mathcal{L}=\sum_{k=0}^{r} c_k \mathcal{L}_k$, with $r\le [d/2]$  have the form $\mathcal{G}_{ab}=0$ where
\begin{equation}  \label{sumform}
\mathcal{G}^a{}_b= \sum_{k=0}^{r} c_k \,
\delta^{a c_1\dots c_k d_1\dots d_k}_{b e_1\dots e_k f_1\dots f_k}\,
R_{c_1d_1} ^{e_1f_1}\dots R_{c_{k}d_{k}}^{e_{k}f_{k}}.
\end{equation}
For the purposes of studying LUV theories, it is also useful to write $\mathcal{G}_{ab}$ in an alternative form which we will make use of below,
\begin{equation}  \label{productform}
\mathcal{G}^a{}_b=\alpha_0\, \delta^{a c_1\dots c_r d_1\dots d_r}_{b e_1\dots e_r f_1\dots f_r}\,
\left(R_{c_1d_1}
{}^{e_1 f_1}+\alpha_1 \delta^{e_1 f_1}_{c_1d_1}\right) \cdots
\left(R_{c_rd_r} {}^{e_r f_r}+\alpha_r
\delta^{e_r f_r}_{c_r d_r}\right).
\end{equation}
The original form of the equations of motion can then be recovered through repeated applications of
the identity
\begin{equation}  \label{factor}
\delta^{a_1\dots a_{p}}_{b_1\dots b_{p}} \delta^{b_{p-1}b_p}_{a_{p-1}a_p} =
2 (D-(p-1))(D-(p-2))\delta^{a_1\dots a_{p-2}}_{b_1\dots b_{p-2}}
\end{equation}
The coefficients $c_k$ are  given by sums of products of the
parameters $\alpha_k$ in (\ref{productform}).  The precise
relation is given in reference \cite{Crisostomo:2000bb}. Inverting
this relation to get the $\alpha_k$'s in terms of the $c_k$'s
requires solving a polynomial equation of order $r$.   Hence the
$\alpha_k$'s are generally complex parameters.

For our purposes, it is convenient to take the parameters
$\alpha_k$ with $k=0,1,\dots,r$ to be real valued and regard the
coefficients $c_k$ in the Lagrangian to be determined by them.  It
is then clear from the expression (\ref{productform}) for
$\mathcal{G}_{ab}$  that if the parameters $\alpha_k$ with
$k=1,\dots,r$ are all distinct, then the theory will have $r$
distinct constant curvature solutions, with Riemann tensors given
respectively by
\begin{equation}
R_{ab}{}^{cd}=-\alpha _{k}\delta _{ab}^{cd}, \qquad k=1,\dots,r
\end{equation}
The LUV
theories discussed above result from setting all the parameters $\alpha _{k}$ with $k=1,\dots r$
equal to a common value $\alpha $. There is then, at least locally, a unique constant curvature vacuum.  If we further set $\alpha =0$, we get a pure Lovelock theory with Lagrangian
$\mathcal{L}=\alpha_0\mathcal{L}_r$, which has flat spacetime as its unique constant curvature vacuum.

The static, spherically symmetric solutions to Lovelock gravity in $D$%
-dimensions can be written in the form
\begin{equation}  \label{schwarz}
ds^2= -f(r)dt^2 + {\frac{dr^2}{f(r)}} + r^2 d\Omega_n^2
\end{equation}
where $n=D-2$. The equations of motion in the form (\ref{factor}) resemble an $r$th order
polynomial equation of order.  It is then not surprising that the metric
function $f(r)$ involves solving an $r$th order algebraic equation \cite%
{Boulware:1985wk}\cite{Wheeler:1985nh}\cite{Wheeler:1985qd} (see also \cite%
{Myers:1988ze} for a detailed analysis of the general case).
We will examine the solutions in Gauss-Bonnet gravity, in which $r=2$, in some detail.  In this case the solutions for the metric function $f(r)$ has two branches, given by
\begin{equation}\label{fpm}
f_\pm (r) = 1- {\frac{r^2}{2}}\left( -(\alpha_1+\alpha_2)\pm\sqrt{%
(\alpha_1-\alpha_2)^2 + {\frac{4}{r_0^4}} \left({\frac{r_0}{r}}\right)^{D-1}}%
\right).
\end{equation}
As $r$ tends to infinity the branches $f_+$ and $f_-$ approach constant curvature metrics with
curvatures $-\alpha_2$ and $-\alpha_1$ respectively.  If we take $\alpha_2=0$, then the solutions in the  $f_+$ brance are asymptotically flat.
Keeping the leading order correction to the flat metric in this case, we have as $r$ tends to infinity
\begin{equation}\label{schwarzlimit}
f_+\simeq 1-{\frac{1}{\alpha_1 r_0^2}}\left({\frac{r_0}{r}}\right)^{D-3}.
\end{equation}
This is the correct fall-off for a black hole of finite ADM mass in $D$-dimensions.  However, the metric differs from the Schwarzschild metric at smaller scales.

We can take the LUV limit of the Gauss-Bonnet theory by setting  $\alpha_1=\alpha_2= \epsilon/l^2$, with $\epsilon=\pm 1$.  The expressions for the metric functions simplify considerably in this limit \cite{Crisostomo:2000bb} to be
\begin{equation}
f_\pm (r) = 1 + \epsilon {r^2\over l^2} \mp  \left( {\frac{r_0}{r}}\right)^{(D-5)\over 2}
\end{equation}
For $D>5$, the branch $f_+$ then gives black holes in (anti-)deSitter, while the branch $f_-$ gives naked singularities. In each case, the falloff to the vacuum is slower than in Schwarzschild-(A)dS, and the ADM mass would therefore be infinite.  However, it was shown in  \cite{Aros:1999kt}\cite{Crisostomo:2000bb} that an appropriate definition of the mass in LUV theories gives a finite result.


One can further take the limit $l\rightarrow\infty$ to get pure Gauss-Bonnet theory.  We then have simply
\begin{equation}
f_\pm (r) = 1  \mp {\frac{|C|}{r^{(D-5)\over 2}}}.
\end{equation}
For $D>5$ the two branches again yield either black holes, or naked curvature singularities.  The case
$D=5$, however, is particularly interesting.  In this case the metric function is simply a constant.
Rescaling the time and radial coordinates, the metric can be
written as
\begin{equation}
ds^{2}=-d\tilde{t}^{2}+d\tilde{r}^{2}+\alpha ^{2}\tilde r^2 d\Omega _{3}^{2}.
\label{missing}
\end{equation}
where $\alpha<1$ for the $f_+$ corresponding to $3$-spheres with
missing solid angle,  while $\alpha>1$ for $f_-$ corresponding to
an excess of solid angle.  In both cases there is a conical
singularity at the origin.   Other solutions to Lovelock theory
with conical singularities at the origin (but with a different
geometry in the constant $(r,t)$ section were obtained recently
\cite{Dehghani}.

Note that $D=5$ is the lowest dimension for which the Gauss-Bonnet term is dynamically relevant.
The situation is similar \cite{Crisostomo:2000bb} for all pure Lovelock theories, $\mathcal{L}=\mathcal{L}_p$, in dimension $D=2p+1$.  The static, spherically symmetric solutions have spheres $S^{2p-1}$ with missing (or excess) solid angle.
The first example
with $p=1$, \textit{i.e.} pure Einstein gravity, in $D=3$ is a familiar
one. Mass in this theory corresponds to point conical defects in an
otherwise flat spacetime.
In the general case spacetime is not flat, but satisfies the  pure Lovelock field
equations.

\section{Black strings and branes in LUV theories}

In pure Einstein gravity, we know that we can construct black brane solutions
by adding flat directions to any  black hole
solution.  More precisely, if we have a spacetime metric $\hat
g_{\mu\nu}$ that solves the vacuum Einstein equations in $D$-dimensions,
then the metric
\begin{equation}  \label{onemore}
ds^2= dz^2 + \hat g_{\mu\nu}dx^\mu dx^\nu
\end{equation}
solves the vacuum field equations in $D+1$ dimensions.
If $\hat g_{\mu\nu}$ is the metric for a black hole with horizon topology $\Sigma$, then the new
metric is a black string with horizon topology $\Sigma\times R^1$.  The procedure can be iterated to generate black branes with any number of flat spatial dimensions tanget to the horizon.

It is straightforward to check that this simple construction does
not work in general Lovelock theories.  In fact, it fails already
in Einstein gravity with a nonzero cosmological constant. Consider
Einstein's equations $R_{ab}-{1\over 2}g_{ab}R+\Lambda g_{ab}=0$
for the $(D+1)$-dimensional metric (\ref{onemore}). Assume that
$\hat g_{\mu\nu}$ satisfies these same equations. Note that all
components of the Riemann tensor of the metric (\ref{onemore})
with one, or more, $z$ indices necessarily vanish.  In particular,
this implies that $R_{\mu\nu}=\hat R_{\mu\nu}$ and $R=\hat R$
where the hatted quantities are those computed from  $\hat
g_{\mu\nu}$.  The $\mu\nu$ components of the field equations are
then identical to those for $\hat g_{\mu\nu}$ which are satisfied
by assumption.  The $zz$ component of the field equations,
however, requires ${1\over 2}\hat R-\Lambda=0$, which conflicts
with the trace of the lower dimensional field equations which
implies ${2-D\over 2}\hat R + D\Lambda = 0$.  We then conclude
that the $(D+1)$-dimensional metric (\ref{onemore}) satisfies the
field equations only for $\Lambda=0$, pure Einstein gravity.

Our first new result is to observe that, while adding flat
directions does not generally work in Lovelock gravity,  it does
work in all pure Lovelock theories, with Lagrangians
$\mathcal{L}=\mathcal{L}_p$.  The result above for pure Einstein
gravity can then be seen as the  particular case $p=1$ of this
more general statement. If $\hat g_{\mu\nu}$ solves the field
equations of the pure Lovelock theory in $D$-dimensions, then the
metric (\ref{onemore}) solves the pure Lovelock field equations in
$D+1$ dimensions. This is easily seen from the field equations
(\ref{sumform}), with only $c_p$ nonzero, in the following way.
Note that as before all the components of the Riemann curvature
tensor with one, or more, $z$ indices necessarilly vanish.  The
field equations $\mathcal{G}^\mu{}_\nu=0$ are then again identical
to the $D$-dimensional field equation
$\hat{\mathcal{G}}^\mu{}_\nu=0$ for $\hat g_{\mu\nu}$ and
hence are satisfied by assumption. The equation $\mathcal{G}^z{}_z=0$ is given
by the trace $\hat\mathcal{G}^\mu{}_\mu$ and hence also vanishes.
Note that, as in the pure Einstein case, the trace of the pure
Lovelock field equations for $\mathcal{L}=\mathcal{L}_p$ is 
equal to $\mathcal{L}_p/\sqrt{-\hat g}$ times a constant.

This result leads to some intriguing possibilities. In pure Einstein theory,
flat directions can be added to the point-like conical defects of $D=3$
gravity to get cosmic strings in $D=4$ and branes in higher dimensions,
which are surrounded by circles with deficit (surfeit) angle. These defects play
important roles in, for example, the Weyl solutions and, in particular, in the C-metric.
There is great interest in higher dimensional generalizations of these spacetimes (see {\it e.g.} \cite{Emparan:2001wk}\cite{Charmousis:2003wm}).
If we take the case of the pure Gauss-Bonnet gravity, $\mathcal{L}_2$, then we
have in $D=5$ a static, spherically symmetric solutions with missing (excess) angle
on an $S^3$. If we add one flat direction, we get a string in $D=6$
surrounded by an $S^3$ with missing (excess) solid angle.  It is tempting to think
that this string might serve as a starting point for finding analogues of the Weyl solutions or the
C-metric in $D=6$.

The next step is to attempt to find a construction of black strings and branes that allows us to add
back in the lower order terms in the Lovelock Lagrangian.
Here, we are able to make some partial progress by again take results from Einstein gravity as a guide.
If one considers Einstein gravity with a negative cosmological
constant $\Lambda =-1/l^{2}$, then we know that if a metric $\hat{g}_{\mu
\nu }$ in $D$ dimensions is Ricci-flat, then the metric
\begin{equation}\label{yetanother}
ds^{2}={\frac{l^{2}}{z^{2}}}(dz^{2}+\hat{g}_{\mu \nu }dx^{\mu }dx^{\nu })
\label{adsstring}
\end{equation}%
solves the field equations in $D+1$ dimensions (see \textit{e.g.} \cite%
{Chamblin:1999by}). If we try this in Lovelock theory, we find the
following result. If $\hat{g}_{\mu \nu }$ solves the field
equations of pure $D$-dimensional Lovelock
gravity with Lagrangian $\mathcal{L}_p$, then the metric (\ref%
{adsstring}) solves the field equations of the $(D+1)$-dimensional
LUV theory,  also with the highest interaction term
$\mathcal{L}_p$, but with $\alpha =1/l^{2}$.

To keep the formulas manageable, we will focus on the case $p=2$
of Gauss-Bonnet gravity.   The proof straightforwardly generalizes
to  higher order LUV theories.  Our $D$-dimensional spacetime
$\hat g_{\mu\nu}$ is then assumed to satisfy the equations of
motion of pure Gauss-Bonnet gravity
\begin{equation}  \label{pureGB}
0= \delta^{\mu \rho_1\dots \rho_{4}}_{\nu\sigma_1\dots \sigma_{4}}\hat
R_{\rho_1\rho_2}{}^{\sigma_1\sigma_2}\hat
R_{\rho_3\rho_4}{}^{\sigma_3\sigma_4}
\end{equation}
where $\hat R_{\mu\nu\rho}{}^\sigma$ is the Riemann tensor of $\hat
g_{\mu\nu}$. If we let $g_{ab}$ denote the metric in (\ref{adsstring}) then
its nonzero curvature components are given by
\begin{equation}  \label{components}
R_{\mu z}{}^{\nu z} = -{\frac{1}{l^2}} \delta_\mu^\nu \qquad
R_{\mu\nu}{}^{\rho\sigma} = {\frac{z^2}{l^2}} \hat R_{\mu\nu}{}^{\rho\sigma}
- {\frac{1}{l^2}} \delta_{\mu\nu}^{\rho\sigma}
\end{equation}
The field equations of the LUV  theory are given by $\mathcal{G}^a{}_b=0$.
After plugging in the curvature components (\ref{components}) the
$\mathcal{G}^{z}{}_z$ component of the field equations  becomes
\begin{equation}
0=\delta _{\sigma _{1}\dots \sigma _{4}}^{\rho _{1}\dots \rho _{4}}\left( {%
\frac{z^{2}}{l^{2}}}\hat{R}_{\rho _{1}\rho _{2}}{}^{\sigma _{1}\sigma
_{2}}+(\alpha -{\frac{1}{l^{2}}})\delta _{\rho _{1}\rho _{2}}^{\sigma
_{1}\sigma _{2}}\right) \left( {\frac{z^{2}}{l^{2}}}\hat{R}_{\rho _{3}\rho
_{4}}{}^{\sigma _{3}\sigma _{4}}+(\alpha -{\frac{1}{l^{2}}})\delta _{\rho
_{3}\rho _{4}}^{\sigma _{3}\sigma _{4}}\right) .
\end{equation}%
Having made the choice $\alpha =1/l^{2}$, we see that  this is simply proportional to the trace
of the equations of motion (\ref{pureGB}) for $\hat{g}_{\mu \nu }$. Plugging into the
$\mathcal{G}^\mu{}_\nu$ components of the field equations gives
\begin{eqnarray}
0 &=&\delta _{\nu \sigma _{1}\dots \sigma _{4}}^{\mu \rho _{1}\dots \rho
_{4}}\left( {\frac{z^{2}}{l^{2}}}\hat{R}_{\rho _{1}\rho _{2}}{}^{\sigma
_{1}\sigma _{2}}+(\alpha -{\frac{1}{l^{2}}})\delta _{\rho _{1}\rho
_{2}}^{\sigma _{1}\sigma _{2}}\right) \left( {\frac{z^{2}}{l^{2}}}\hat{R}%
_{\rho _{3}\rho _{4}}{}^{\sigma _{3}\sigma _{4}}+(\alpha -{\frac{1}{l^{2}}}%
)\delta _{\rho _{3}\rho _{4}}^{\sigma _{3}\sigma _{4}}\right)  \\
&&+8\delta _{\nu \sigma _{1}\dots \sigma _{3}}^{\mu \rho _{1}\dots \rho
_{3}}\left( (\alpha -{\frac{1}{l^{2}}})\delta _{\rho _{1}}^{\sigma
_{1}}\right) \left( {\frac{z^{2}}{l^{2}}}\hat{R}_{\rho _{2}\rho
_{3}}{}^{\sigma _{2}\sigma _{3}}+(\alpha -{\frac{1}{l^{2}}})\delta _{\rho
_{2}\rho _{3}}^{\sigma _{2}\sigma _{3}}\right)   \nonumber
\end{eqnarray}%
For $\alpha =1/l^{2}$ the first term again reduces to the $D$-dimensional equations of
motion for $\hat{g}_{\mu \nu }$, while the second term simply
vanishes. The equations $\mathcal{G}^z{}_\mu=0$ are trivially satisfied, and we have thus shown that adding the construction (\ref{adsstring}) of new solutions holds for LUV theories of maximum order $p=2$ starting with solutions of
pure pure Gauss-Bonnet theory.
The structure of our derivation makes it clear that the result holds for higher order LUV theories as well.

We now prove a related, but more general result as well. Assume now that the
$D$-dimensional metric $\hat{g}%
_{\mu \nu }$ solves the equations of motion of LUV  theory with maximum
interaction $\mathcal{L}_{2}$ and curvature parameter $\beta $,
\begin{equation}\label{original}
0=\delta _{\nu \sigma _{1}\dots \sigma _{4}}^{\mu \rho _{1}\dots \rho
_{4}}\left( \hat{R}_{\rho _{1}\rho _{2}}{}^{\sigma _{1}\sigma _{2}}+\beta
\delta _{\rho _{1}\rho _{2}}^{\sigma _{1}\sigma _{2}}\right) \left( \hat{R}%
_{\rho _{3}\rho _{4}}{}^{\sigma _{3}\sigma _{4}}+\beta \delta _{\rho
_{3}\rho _{4}}^{\sigma _{3}\sigma _{4}}\right) .
\end{equation}%
Now consider a $(D+1)$-dimensional metric of the form
\begin{equation}
ds^{2}={\frac{1}{f(z)^{2}}}(dz^{2}+\hat{g}_{\mu \nu }dx^{\mu }dx^{\nu })
\label{fmetric}
\end{equation}%
and ask if it can solve the field equations of the same LUV  theory, but
with vacuum curvature parameter $\alpha $ instead.  Our previous result
corresponds to the special case that the curvature parameter of the $D$-dimensional metric vanishes.
The curvature of the $(D+1)$-dimensional metric (\ref{fmetric}) is given by
\begin{equation}
R_{\mu z}{}^{\nu z}=(f\partial _{z}^{2}f-(\partial _{z}f)^{2})\delta _{\mu
}^{\nu }\qquad R_{\mu \nu }{}^{\rho \sigma }=f^{2}\hat{R}_{\mu \nu }{}^{\rho
\sigma }-(\partial _{z}f)^{2}\delta _{\mu \nu }^{\rho \sigma }
\end{equation}%
The field equations for the $(D+1)$-dimensional metric (\ref{fmetric}) are again
given by $\mathcal{G}^{a}{}_b=0$. Upon plugging in the components of the Riemann
tensor, the equation $\mathcal{G}^{z}{}_z=0$ becomes
\begin{equation}
0=\delta _{\sigma _{1}\dots \sigma _{4}}^{\rho _{1}\dots \rho _{4}}\left(
f^{2}\hat{R}_{\rho _{1}\rho _{2}}{}^{\sigma _{1}\sigma _{2}}+(\alpha
-(\partial _{z}f)^{2})\delta _{\rho _{1}\rho _{2}}^{\sigma _{1}\sigma
_{2}}\right) \left( f^{2}\hat{R}_{\rho _{3}\rho _{4}}{}^{\sigma _{3}\sigma
_{4}}+(\alpha -(\partial _{z}f)^{2})\delta _{\rho _{3}\rho _{4}}^{\sigma
_{3}\sigma _{4}}\right) .
\end{equation}%
This will be proportional to the trace of the field equations (\ref{original}) for the $D$-dimensional
metric $\hat g_{\mu\nu}$ if the function $f(z)$ satisfies the relation
\begin{equation}
\alpha -(\partial _{z}f)^{2}=\beta f^{2}  \label{fequation}
\end{equation}%
Plugging in the components of the Riemann tensor into the equations $\mathcal{G}^{\mu }{}_\nu=0$ gives the equation
\begin{eqnarray}
0 &=&\delta _{\nu \sigma _{1}\dots \sigma _{4}}^{\mu \rho _{1}\dots \rho
_{4}}\left( f^{2}\hat{R}_{\rho _{1}\rho _{2}}{}^{\sigma _{1}\sigma
_{2}}+(\alpha -(\partial _{z}f)^{2})\delta _{\rho _{1}\rho _{2}}^{\sigma
_{1}\sigma _{2}}\right) \left( f^{2}\hat{R}_{\rho _{3}\rho _{4}}{}^{\sigma
_{3}\sigma _{4}}+(\alpha -(\partial _{z}f)^{2})\delta _{\rho _{3}\rho
_{4}}^{\sigma _{3}\sigma _{4}}\right)  \\
&&+8\delta _{\nu \sigma _{1}\dots \sigma _{3}}^{\mu \rho _{1}\dots \rho
_{3}}\left( (\alpha -(\partial _{z}f)^{2}+f\partial _{z}^{2}f)\delta _{\rho
_{1}}^{\sigma _{1}}\right) \left( f^{2}\hat{R}_{\rho _{2}\rho
_{3}}{}^{\sigma _{2}\sigma _{3}}+(\alpha -(\partial _{z}f)^{2})\delta _{\rho
_{2}\rho _{3}}^{\sigma _{2}\sigma _{3}}\right)   \nonumber
\end{eqnarray}%
The first of terms will vanish by virtue of the equations of motion (\ref{original}) for
the $D$-dimensional metric
$\hat{g}_{\mu \nu }$ if the relation (\ref{fequation}) is satisfied. The second term
will vanish, if the function $f(x)$ also satisfies the additional relation
\begin{equation}
\alpha -(\partial _{z}f)^{2}+f\partial _{z}^{2}f=0.
\end{equation}%
If we write the $D$ and $(D+1)$-dimensional curvature parameters as $\beta =1/l^{\prime 2}$ and $\alpha =1/l^{2}$ respectively, then both relations are satisfied if the conformal factor $f(z)$ in the
$(D+1)$-dimensional metric (\ref{yetanother}) is taken to be
\begin{equation}
f(z)={\frac{l^{\prime }}{l}}\sin (z/l^{\prime }).  \label{function}
\end{equation}%
Note that we can take the $l^\prime\rightarrow\infty$ limit of this result in which the curvature parameter
$\beta=1/l^{\prime 2}$ of the $D$-dimensional metric vanishes.  Our previous result for this case,  that $f(z)=z/l$, is then seen as the small angle approximation to the more general result above.  We also note that, for the sake of definiteness, we have focused on LUV theories with AdS vacua.  However, the above result is only simply modified for LUV theories with deSitter vacua.  If we take $\alpha=-1/l^2$ and
$\beta= -1/l^{\prime 2}$ then we find that the conformal scaling function $f(z)$ must be given by
\begin{equation}
f={\frac{l^{\prime }}{l}}\sinh (z/l^{\prime }).  \label{function 2}
\end{equation}

\section{Concluding Remarks}
We have shown that the construction of black string and brane solutions in pure Lovelock gravity theories and in LUV theories parallels the simple constructions in pure Einstein theory and in Einstein theory with a nonzero cosmological constant respectively.
While this does not fully solve the problem of constructing black brane solutions in Lovelock gravity theories, it represents useful progress in a number of ways.

On one hand, it has been argued in reference \cite{Crisostomo:2000bb} that LUV theories represent the most physically interesting class of Lovelock gravity theories.  From this perspective, our results cover precisely this most interesting case. Further, the simplicity of our results, and their parallels with Einstein gravity, suggest that there may well be other methods of constructing solutions in Einstein gravity that have simple generalizations to pure Lovelock and LUV theories.  In particular, as we commented above, the appearance of string solutions with deficit solid angle on odd dimensional spheres, suggests that we might be able to find analogues of the Weyl solutions and the C-metric in certain of these theories.

On the other hand, one's primary interest may be in corrections to solutions to pure Einstein gravity due to higher curvature Lovelock terms.  In this case, our results are limited, but provide some clues towards finding black string and brane solutions in the general case.  Consider the black hole solutions of Gauss-Bonnet gravity which have the form (\ref{schwarz}) with the metric function $f(z)$ given in equation (\ref{fpm}).  Take $\alpha_2=0$, so that we have an asymptotically flat solution.  For large $r$ this  approaches Schwarzschild as in equation (\ref{schwarzlimit}).  As $r\rightarrow 0$, however, where the curvature becomes large, the solution matches onto a solution to pure Gauss-Bonnet gravity.  If we want to construct a black string solution, in this case, it should asymptote for large $r$ to Schwarzschild with an extra flat direction, the black string of pure Einstein gravity.   Our results suggest that it should also look simple at small radii.  There it should approach  the pure Gauss-Bonnet limit of (\ref{fpm}) again with a flat direction added.  For intermediate radii, the solution will presumably depend on the coordinate along the string in a non-trivial way.



\begin{acknowledgements}
The authors would like to thank the organizers of the \textsl{Scanning New Horizons: GR Beyond 4 Dimensions} workshop at the KITP where this work was carried out.  DK would like to thank Fay Dowker, Satyanarayan Mohapatra and Jennie Traschen for helpful discussions on this subject.
This research was supported in part by the National Science Foundation under Grant No. PHY99-07949 and Grant No. PHYÐ0244801
and by the Natural Sciences \& Engineering Research Council of Canada.
\end{acknowledgements}

\end{document}